\documentstyle[12pt]{article}
\begin{document}

\begin{center}
{\large {\bf A Semiclassical ANEC Constraint on
Classical Traversable Lorentzian Wormholes }}\\[0pt]

\vspace{8mm}

Kamal K. Nandi$^{a,c,}$\footnote{E-mail: kamalnandi@hotmail.com},
Yuan-Zhong Zhang$^{b,c,}$\footnote{E-mail: yzhang@itp.ac.cn}
 and Nail G. Migranov$^{d,}$\footnote{E-mail: migranov@bspu.ru}

\vspace{5mm}

{\footnotesize {\it
 $^a$ Department of Mathematics, University of North Bengal,
       Darjeeling (W.B.) 734 430, India \\
 $^b$ CCAST (World Laboratory), P.O.Box 8730, Beijing 100080, China \\
 $^c$ Institute of Theoretical Physics, Chinese Academy of Sciences, P.O.Box
       2735, Beijing 100080, China \\
 $^d$ DJoint Research Center for Mathematics and Physics (JRCMP),
       Bashkir State Pedagogical University, 3-A,
      October Revolution Street, Ufa 450000, Russia. }}

\end{center}

\vspace{20mm}

\begin{abstract}
The present article lies at the interface between gravity, a
highly nonlinear phenomenon, and quantum field theory. The
nonlinear field equations of Einstein permit the theoretical
existence of classical wormholes. One of the fundamental questions
relates to the practical viability of such wormholes. One way to
answer this question is to assess if the total \emph{volume} of
exotic matter needed to maintain the wormhole is finite. Using
this value as the lower bound, we propose a modified semiclassical
volume Averaged Null Energy Condition (ANEC) constraint as a
method of discarding many solutions as being possible
self-consistent wormhole solutions of semiclassical gravity. The
proposed constraint is consistent with known results. It turns out
that a class of Morris-Thorne wormholes can be ruled out on the
basis of this constraint.

\bigskip

\noindent PACS number(s): 04.70.Dy,04.62.+v,11.10.Kk
\end{abstract}

\vspace{20mm}

 The static, spherically symmetric traversable Lorentzian wormholes
are special classes of theoretical solutions of the highly
nonlinear gravitational field equations of Einstein. The solutions
can be interpreted as objects connecting ("handle") two distant
regions of spacetime. These objects (i.e., wormholes) are threaded
by what is called "exotic" matter. The notion of such has found a
justification also in the wider context of cosmology where one
deals with dark matter or classical phantom field. However, this
article is concerned with a local problem described by static,
spherically symmetric wormhole spacetimes in which it is known
that several pointwise and average energy conditions are violated.
We shall discuss here the violation of the weakest of all energy
conditions, viz., the Averged Null Energy Condition (ANEC), and
refer to the ANEC violating matter as exotic. An immediate
question to be asked is this: How to quantify the total amount of
exotic matter present in a wormhole spacetime? The question has
significant relevance to some global results such as the
singularity (see, e.g. [1-3]) or positive mass theorems of general
relativity [4]. Normally, the ANEC is stated as an integral of the
stress tensor averaged along a complete null geodesic being
non-negative. This is a line integral with dimensions
(mass)/(area), and hence is not very useful in providing
information about the {\it total }amount of exotic matter.
Considering this, Visser, Kar and Dadhich [5] proposed a volume
integral quantifier that has been properly modified recently [6]
on physical grounds. We are going to use this in what follows and
consider traversable wormholes joining two asymptotically flat
regions.

The object of the present paper is to propose a constraint on the
ANEC violating matter in the form of an inequality in which the
classical ANEC volume integral appears as the lower bound to the
generalized ANEC integral following from Quantum Field Theory.
That is the reason why we termed our constraint as semiclassical
and its importance lies in the fact that it could be used to test
the physical viability of any given static spherically symmetric
wormhole. (We call any wormhole physically viable if the classical
ANEC violation is \emph{finite}. This finiteness is the primary
condition for Morris-Thorne type of wormholes to be threaded by
any quantum field. For simplicity, we consider here only the
quantum Klein-Gordon field.) We show that a well known class of
Morris-Thorne wormholes is not quantum mechanically viable.

As a first step, we shall provide the classical volume ANEC
integral. As the next step, we shall consider the generalized
quantum ANEC suggested by Yurtsever [7]. He concluded that the
Klein-Gordon stress energy of semiclassical gravity can support
wormholes that are only roughly of the Planck size. His arguments
are based on the convergence of line integrals but we will see
that the volume ANEC integral does not always preserve this
convergence. In the final step, we apply the constraint to
Morris-Thorne wormholes.

Let us consider the wormhole solution in the general form%
\begin{equation}
ds^{2}=-\exp [2\Phi (l)]dt^{2}+dl^{2}+r^{2}(l)\left( {d\theta
^{2}+\sin ^{2}\theta d\varphi ^{2}}\right) .
\end{equation}%
The throat of the wormhole occurs at $l=0.$ The volume ANEC
integral to be calculated in an orthonormal frame is [6] (We take
$G=c=\hbar =1$):
\begin{equation}
\Omega _{ANEC}=\int\limits_{x_{th}}^{\infty }\int \int [T_{\mu \nu
}k^{\mu }k^{\nu }]\sqrt{-g_{4}}d^{3}x
\end{equation}%
for null $k^{\mu }$ and $g_{4}=\det \left\vert g_{\mu \nu
}\right\vert $. One might notice that the volume measure is just
the one appearing in the Tolman-Komar integral [8] with the
difference that the radial integration is from the throat to
$\infty $ for one side of the wormhole. The integral resembles the
usual definition of quasi-local energy. The quantity $T_{\mu \nu
}k^{\mu }k^{\nu }$ is a general covariant scalar but depends on
the congruence {\it C }of null geodesics filling the entire region
of space [9]. For further details and physical justification of
Eq.(2), see Ref.[6]. The energy and pressure densities in that
region as measured in the local
orthonormal Lorentz frames (\symbol{94}) for the metric (1) are:%
\begin{equation}
\rho =T_{\widehat{t}\widehat{t}}=-\frac{2r^{\prime \prime }}{r}+\frac{%
1-r^{\prime 2}}{r^{2}}
\end{equation}%
\begin{equation}
p_{l}=T_{\widehat{l}\widehat{l}}=\frac{2\Phi ^{\prime }r^{\prime }}{r}-\frac{%
1-r^{\prime 2}}{r^{2}}
\end{equation}

\begin{equation}
p_{\theta }=T_{\widehat{\theta }\widehat{\theta }}=p_{\varphi }=T_{\widehat{%
\varphi }\widehat{\varphi }}=\frac{1}{2}[\Phi ^{\prime \prime
}+(\Phi ^{\prime })^{2}+\frac{\Phi ^{\prime }r^{\prime }+r^{\prime
\prime }}{r}]
\end{equation}%
where $X^{\prime }\equiv dX/dl$. It is argued in Ref.[5] that the
transverse components are associated with \textquotedblleft
normal\textquotedblright\ matter and only the remaining components
are to be considered for
investigating the volume ANEC violation. Therefore, Eq.(2) translates into%
\begin{equation}
\Omega _{ANEC}=\frac{1}{2}\times \int\limits_{-\infty }^{+\infty
}{(\rho +p_{l})e}^{\Phi }r^{2}dl.
\end{equation}%
There is a factor of $4\pi $ multiplying the integral coming from the $%
\theta ,\varphi $ integration, but $\rho $ and $p_{l}$ each has
$(1/8\pi )$ as factors. Hence the resulting factor of
$(\frac{1}{2})$ that actually cancels out when we compute $\Omega
_{ANEC}$ for two mouths of the symmetric wormhole. We say that
ANEC is satisfied if $\Omega _{ANEC}$ is non-negative,
but for classical wormhole spacetime, ANEC is always violated, that is, $%
\Omega _{ANEC}$ is negative. We advocate using the volume integral
(6) (multiplied by 2) for assessing the total amount of ANEC
violating matter in preference to the conventional line integral,
viz.,
\begin{equation}
V=\frac{1}{8\pi }\int_{\gamma }T_{\mu \nu }k^{\mu }k^{\nu
}dv=-\frac{1}{4\pi }\int_{-\infty }^{+\infty }e^{-\Phi }\left(
\frac{r^{\prime }}{r}\right) ^{2}dl
\end{equation}%
where $dv=e^{\Phi }dl$, $v$ being the affine parameter along the
null geodesic $\gamma $.

The next step is to consider the quantum picture, that is, the
quantum field theory in curved spacetime, or semiclassical
relativity: $G_{\mu \nu }=8\pi \left\langle T_{\mu \nu
}\right\rangle $. The situation here is far more complex than the
classical picture:\ No complete characterization of the stress
tensor in the semiclassical Einstein equations is available as
yet. Early works [10-12] have shown that, under certain asymptotic
regularity conditions, the ANEC is satisfied by minimally coupled
scalar field in four dimensional Minkowski spacetime and
conformally coupled field in the curved two-dimensional spacetime
in all quantum states forming a subset of the standard Fock space.
These results have been strengthened by an analysis based on an
algebraic approach devised by Wald and Yurtsever [13]. If the null
geodesic is achronal, then the ANEC is satisfied when the Casimir
vacuum contribution is subtracted from the stress energy resulting
into Ford-Roman difference inequalities [14]. Yurtsever has
provided a proof of this inequality in globally hyperbolic
two-dimensional spacetimes [15].

However, crucial for our analysis is the suggestion of a
generalized ANEC by Yurtsever [7] which might hold, unlike the
conventional quantum ANEC, in a
four-dimensional curved spacetime given by%
\begin{equation}
\beta (k)=\mathop {\inf }\limits_\omega
\int_{\gamma }\left\langle
\omega \left\vert T_{\mu \nu }\right\vert \omega \right\rangle
k^{\mu }k^{\nu }dv.
\end{equation}%
The quantum stress tensor $\left\langle T_{\mu \nu }\right\rangle
$ satisfies the generalized ANEC along a null geodesic $\gamma $
if the one-form $\beta (k)>-\infty $. The infimum is taken over
all Hadamard states
$\omega $ of the quantum field and the tangent vector is defined by $%
k^{\alpha }=\frac{d\gamma ^{\alpha }}{dv}$. The integral on the
left is
further refined into $\beta _{c}(k)$ by introducing a weighting function $%
c(x)$ but, with Yurtsever [7], we assume that $\beta
_{c}(k)=[c(0)]^{2}\beta (k)$. The value of $\beta (k)$ can be
obtained by using the scaling argument. Hereafter, we shall
closely follow the arguments in Ref.[7]. Given any arbitrary
four-dimensional spacetime ({\bf M},$g$) in which the massless
Klein-Gordon field $\phi $ satisfies generalized ANEC along the
null geodesic $\gamma $, the scaling argument requires us to
consider a new spacetime ({\bf M},$\kappa ^{2}g$) where $\kappa
>0$. The renormalization procedure (See Refs.[7,16]) involving the
two-point function contributes, apart from the simply scaled term
$\kappa ^{2}\left\langle \omega \left\vert T_{\mu \nu }\right\vert
\omega \right\rangle $, two additional terms to the value of
$\left\langle \omega \left\vert T_{\mu \nu }\right\vert \omega
\right\rangle $ that are of the form $[a^{(1)}H_{\mu \nu
}+b^{(2)}H_{\mu \nu }]$ $\kappa ^{-2}\ln \kappa $ where $a$ and
$b$ are dimensionless constants having values of the order of
$10^{-4}$ in Planck units,
\begin{equation}
^{(1)}H_{\mu \nu }\equiv 2R_{;\mu \nu }+2RR_{\mu \nu }-g_{\mu \nu
}(2\Box R+\frac{1}{2}R^{2})
\end{equation}%
\begin{equation}
^{(2)}H_{\mu \nu }\equiv R_{;\mu \nu }-\Box R_{\mu \nu }+2R_{\mu
}^{\alpha }R_{\alpha \nu }-\frac{1}{2}g_{\mu \nu }(\Box R+R^{\eta
\delta }R_{\eta \delta }).
\end{equation}%
For a general spacetime, $\beta (k)$ scales as [7]%
\begin{equation}
\overline{\beta (k)}=\frac{1}{\kappa ^{3}}\beta (k)+\frac{\ln \kappa }{%
\kappa ^{3}}\int_{\gamma }(a^{(1)}H_{\mu \nu }k^{\mu }k^{\nu
}+b^{(2)}H_{\mu \nu }k^{\mu }k^{\nu })dv.
\end{equation}%
We shall replace the line integral measure above by the volume
integral measure as in (2) such that $\overline{\beta (k)}$
changes
to%
\begin{equation}
\overline{\beta_{1} (k)}=\frac{1}{\kappa}\beta_{0} (k)+\frac{\ln \kappa }{%
\kappa}\int_{-\infty }^{\infty }(a^{(1)}H_{\mu \nu }k^{\mu }k^{\nu
}+b^{(2)}H_{\mu \nu }k^{\mu }k^{\nu }){e}^{\Phi }r^{2}dl
\end{equation}%
for the spacetime (1) where $\beta_{0} (k)$ is the volumized
version of Eq.(8). The expressions for $^{(1)}H_{\mu \nu }k^{\mu
}k^{\nu }$ and $^{(2)}H_{\mu \nu }k^{\mu }k^{\nu }$ have been
computed by Yurtsever [7]. Assuming that the same null congruences
fill the scaled and unscaled
spacetimes, we shall compare the value of $\overline{\beta_{1} (k)}$ with $%
\overline{\Omega }_{ANEC}$ which is just the scaled value of $\Omega _{ANEC}$%
. If the ANEC\ violating matter is to be supportable by the
renormalized quantum stress tensor, then, we conjecture, on
dimensional grounds, that the following inequality
\begin{equation}
\overline{\beta_{1} (k)}\leq \overline{\Omega }_{ANEC}\Rightarrow
\left\vert \overline{\beta_{1} (k)}\right\vert \geq \left\vert \overline{%
\Omega }_{ANEC}\right\vert
\end{equation}%
be satisfied. That is, the classical quantity $\left\vert \overline{\Omega }%
_{ANEC}\right\vert $ plays the role of a finite lower bound to
$\left\vert \overline{\beta_{1} (k)}\right\vert $. In this sense,
the inequality (13) may be regarded as a modification to the
generalized ANEC. In order that the conjectured inequality makes
sense, it is
necessary that $%
\left\vert \overline{\Omega }_{ANEC}\right\vert <\infty $ be
negative but finite. Any classical asymptotically flat traversable
wormhole satisfying this condition is defined in this paper as
physically viable. The constraint (13) is not only nontrivial, as
the latter arguments will show, but is also of sufficiently
general nature in that we can apply it to any given spherically
symmetric wormhole spacetime.

With Eqs.(6) and (12) at hand, let us consider the general form of
a
near-Schwarzschild traversable wormhole of mass $M$, asymptotically $%
(l\rightarrow \pm \infty )$ described by the metric functions%
\begin{equation}
r(l)\simeq \left\vert l\right\vert -M\ln \left( \left\vert
l\right\vert /r_{0}\right) ,\Phi (l)\simeq -M/\left\vert
l\right\vert
\end{equation}%
where $r_{0}\sim 2M$ is the throat radius. To proceed with the
integration (6) using Eqs.(3) and (4), we note that $r(l)\simeq
\left\vert l\right\vert [1-(M/\left\vert l\right\vert )\ln \left(
\left\vert l\right\vert
/r_{0}\right) ]\simeq l$ without much error since the functions $%
1/\left\vert l\right\vert $ and $\ln \left\vert l\right\vert
/\left\vert l\right\vert $ almost compensate each other in the
intervals $\left\vert
l\right\vert \geq 2M.$ Using this fact, and computing $r^{\prime }$, $%
r^{\prime \prime }$ etc we can integrate Eq.(7) over $\left\vert
l\right\vert \geq 2M$ to find that $V\simeq -\frac{0.17}{\pi M}.$
Under the same approximation, we find from Eq.(6) that
\begin{equation}
\Omega _{ANEC}\simeq -0.39M
\end{equation}%
and hence ANEC is violated. An examination of the integrand in
Eq.(6) reveals that, under the scaling $\overline{g}=\kappa
^{2}g$, ($\kappa >0$),
we have $\overline{r}_{0}=\kappa r_{0}$, and consistent with this, $%
\overline{\Omega }_{ANEC}=\kappa \Omega _{ANEC}$. This immediately gives $%
\overline{M}=\kappa M$. Thus, $\left\vert \overline{\Omega }%
_{ANEC}\right\vert <\infty $ showing that the wormhole could be
supported by a quantum scalar field. Using the expressions given
in Ref.[7], we find that
the extra renormalization contributions work out to finite negative values%
\begin{equation}
\int_{-\infty }^{+\infty }{}^{(1)}H_{\mu \nu }k^{\mu }k^{\nu
}{e}^{\Phi }r^{2}dl\simeq -\frac{23.29}{M}
\end{equation}%
\begin{equation}
\int_{-\infty }^{+\infty }{}^{(2)}H_{\mu \nu }k^{\mu }k^{\nu
}{e}^{\Phi }r^{2}dl\simeq -\frac{9.04}{M}.
\end{equation}%
Let $B(M)$ and $B(\overline{M})$ denote respectively the values of
$\beta (k) $ and $\overline{\beta (k)}$ such that we can rewrite
Eq.(12) as
\begin{equation}
B(\kappa M)=\frac{1}{\kappa ^{3}}B(M)+\frac{\ln \kappa }{\kappa ^{3}}%
(10^{-3}c/M)
\end{equation}%
where the numerical constant $\left\vert c\right\vert \sim 1$.
Assuming that
$\left\vert B(1)\right\vert \sim 1$ for a Planck mass $M\sim 1$, we have, $%
\overline{M}=\kappa $ and
\begin{equation}
B(\overline{M})\simeq \left( \frac{1}{\overline{M}^{3}}\right)
[c_{1}+10^{-3}c_{2}\ln \overline{M}]
\end{equation}%
where, again, $\left\vert c_{1}\right\vert \sim \left\vert
c_{2}\right\vert \sim 1$. For reasonable values of $\overline{M}$,
the logarithmic term in the square bracket can be ignored compared
to the first term and so, dropping bars, we are left with
$\left\vert B(M)\right\vert \simeq 1/M^{3}$. Using the condition
(13), viz., $\left\vert B(M)\right\vert \geq \left\vert \Omega
_{ANEC}\right\vert $, and putting in the corresponding values, we
see that $M^{4}\leq c_{3}$ where $\left\vert c_{3}\right\vert \sim
1$. That is, a wormhole to be supportable by a massless quantum
Klein-Gordon field must only be of the order of a Planck mass. We
see that the volume integral approach also supports the Planck
size constraint in the case of near Schwarzschild wormholes
imposed by the earlier consideration of line integrals in the form
of the constraint $\left\vert \beta (k)\right\vert \geq \left\vert
 {V}\right\vert $. A similar constraint on size is provided also by
the Ford-Roman Quantum Inequality (FRQI) stated in the form
somewhat similar to the \textquotedblleft
energy-time\textquotedblright\ inequality [17]. The FRQI has been
recently discussed in connection with several classical wormhole
solutions from the minimally coupled theory [18].

The modification suggested in (13) is not trivial. The difference
is that the volume integral (6) and the line integral (7) do
\emph{not} necessarily lead to similar results in the case of
solutions \emph{deviating} from the near-Schwarzschild metric
(14). To illustrate our point, we consider first the case where
the two integrals (line and volume ANEC) do lead to finite values.
This is given, for instance, by the widely discussed class of
\textquotedblleft zero tidal force\textquotedblright\ traversable
wormholes
[19]. One typical member is given by%
\begin{equation}
\Phi (l)=0,r(l)=\sqrt{l^{2}+b^{2}}.
\end{equation}%
The throat occurs at $l=0$, or at $r=b$ where the parameter $b>0$.
It immediately follows that the integrals (6) and (7) are finite.
In particular, $\Omega _{ANEC}=-\frac{\pi b}{2}$, and thus ANEC is
violated. Also, the integrals (16) and (17) easily work out to
finite values that are of the order of $\left\vert c_{4}/b\right\vert $ where $%
\left\vert c_{4}\right\vert \sim 1.$ The inequality (13) is
satisfied only if $b^{2}\leq 1$, that is, the radius of the
wormhole is of the Planck size and conversely. It is clearly a
physically viable wormhole. Similar considerations apply to many
other kinds of wormholes [20] with different expressions for
$r(l)$ for which (6) and (12) converge. All these wormholes are
physically realistic.

Consider next another class of Morris-Thorne solutions given by [17,21]%
\begin{equation}
\Phi (l)=0,r(l)\simeq \left\vert l\right\vert -M\ln \left(
\left\vert l\right\vert /r_{0}\right) .
\end{equation}%
For this, the line integral of Eq.(7) for $V$ \emph{converges} to
a finite negative value. Assuming that the GANEC holds, the
arguments surrounding Eqs.(18) and (19) (but without the
consideration of the volume integral) would lead to the conclusion
that these wormholes are also of Planck size and physically
realistic. However, this is not necessarily the case. The $\Omega
_{ANEC}$ of Eq.(6) [and hence $\overline{\Omega }_{ANEC}$]
produces a logarithmic {\it divergence} in the asymptotic region
since $\rho +p_{l}\sim O(l^{-3})$ indicating that any quantum
field is unlikely to support this huge quantity of exotic matter.
The divergence on the left hand side renders the inequality (13)
physically meaningless. The example clearly illustrates that the
conclusions based on the volume integrals could be radically
different from those based on line integrals. Wormholes of the
type Eq.(21) may thus be physically unrealistic.

Eq.(6) provides a new classical volume ANEC in the form $\Omega
_{ANEC}\geq 0$ (or, which is the same, $\overline{\Omega }%
_{ANEC}\geq 0$) with similar volume measures adopted for $\overline{%
\beta_{1} (k)}$ in the scaled spacetime. The constraint imposed by
the quantum
field theory on $\Omega _{ANEC}$ is that it must be negative and{\it \ }%
finite, but not too large (not more than $\sim 10^{4}M$) knowing that $%
\overline{\beta_{1} (k)}$ is finite for asymptotically flat
wormholes. The finiteness of $\Omega _{ANEC}$ is guaranteed via
the local classical conservation law $T_{\mu ;\nu }^{\nu }=0 $
that describes the exchange of energy between (exotic) matter and
gravitation together with the fall-off $\rho \sim O(l^{-4})$. In
the geodesic orthonormal coordinates, the law leads to a
conserved quantity $P\mu =\int T_{\widehat{0}\widehat{\mu }}\sqrt{-g_{4}}%
d^{3}x$, and we obtain finite values for total energy
$P_{0}$.(Note that $P\mu $ is similar to $\Omega _{ANEC}$ except
in the radial integration limits). Note further that stress
tensors of well known classical fields (like in the minimally or
conformally coupled theories) satisfy $T_{\mu ;\nu }^{\nu }(\phi
)=0$ independently of the Bianchi identities $G_{\mu ;\nu }^{\nu
}\equiv 0$ together with the desired fall-off for general
spherically symmetric configurations. However, for
\emph{arbitrary} choice of metric functions, one can compute from
the Einstein tensor some expressions for $T_{\mu }^{\nu }$, and
that $T_{\mu ;\nu }^{\nu }=0$ follows {\it only} as a result of
the Bianchi identities $G_{\mu ;\nu }^{\nu }\equiv 0$. It is not
guaranteed that these local conservation laws, in turn, would
provide the desired decay law ($\rho +p_{l})\sim O(l^{-4})$. An
illustration is provided by the example in (21), for which ($\rho
+p_{l})\sim O(l^{-3})$. Although there is asymptotic decay, it is
quite unlikely that the wormhole is threaded by a finite quantity
of ANEC violating matter.

Conversely, in the context of quantum field theory, an interesting
example is this: The natural in-vacuum states of any scalar field
in the Minkowski spacetime will have a Casimir energy density
$\rho =$ negative constant $=-a$ (say) over all space. This
essentially represents the vacuum solution of semi-classical
relativity, having no classical gravity counterpart. (Note
parenthetically that the \textquotedblleft Casimir
vacuum\textquotedblright\ Morris-Thorne-Yurtsever quantum
wormholes require a plate separation smaller than the electron
Compton wavelength [22].) Nonetheless, a na\~{\i}ve integration in
(6) over the Minkowski space does give $\Omega _{ANEC}=-\infty $.
Now, it is understood that there is no unique way of making the
transition from classical to quantum regime. The quantum system
may contain interesting aspects of the true situation which
disappear in the correspondence-principle limit. It is not clear
if there is any classical curved spacetime counterpart to the
Minkowski spacetime quantum field theory. Clearly, the
semiclassical ANEC constraint can not be meaningfully applied in
this case. In a related context, it might be of interest to note
that Popov [23] has obtained, under a \textquotedblleft
subtraction\textquotedblright\ scheme, an analytic approximation
of the stress energy tensor of quantized \emph{massive}
scalar fields in static spherically symmetric spacetimes with topology {\it R%
}$^{2}\times ${\it S}$^{2}$. The stress tensor supports a
Morris-Thorne wormhole if the curvature coupling parameter $\xi
\geq 0.2538$.

Recall that the situation is different in the near-Schwarzschild
wormhole case. The massless quantum Klein-Gordon field does have a
classical analogue in the sense that the wormhole Eq.(14) {\it is}
an approximation to a physically meaningful wormhole solution in
the minimally coupled field
theory with a {\it negative} kinetic term in the Lagrangian, viz., $%
L_{matter}=-(1/8\pi )\partial _{\mu }\phi \partial ^{\mu }\phi $
[24]. (This result is important as the explicit existence of a
classical scalar field is useful for the regularization program
and the correspondence limit.)
Interestingly, the example in Eq.(20) too describes an exact {\it extremal }%
zero total mass{\it \ }wormhole solution in that theory [20]. It
has been shown that zero mass wormholes, including slightly
massive ones, are stable [25]. Physically realistic stable
wormholes following from Hilbert-Einstein action with a well
defined scalar field matter Lagrangian such as above have $\Omega
_{ANEC}<-\infty $. Microscopic quantum wormholes also require this
classical condition to hold and stability of such wormholes lends
support to the guess (not a proof) that the condition could also
be a key ingredient in a general classical stability analysis. We
do not argue here that macroscopic wormholes can not occur in
nature just because of the stability criterion. In fact, if one
includes classical matter field $T_{\mu \nu }^{C}$ in addition to
quantum field $T_{\mu \nu }^{Q}$, one could have macroscopic
wormholes supported by quantum field [26]. What we do argue is
that the validity of the
generalized ANEC in the entire spacetime giving a finite $\overline{\beta_{1} (k)%
}$ requires that $\Omega _{ANEC}<-\infty $, and this condition is
sufficient to rule out many classical macroscopic configurations.
The requirement of a correspondence limit could be an additional
condition on the quantum scenario, but we do not emphasize it.

To summarize, we saw, by employing the volume integrals, that the
scaling argument holds and that classical ANEC violation can be
supported by semiclassical gravity if the wormhole is microscopic.
However, for non-Schwarzschild $\Phi =0$ wormholes that abound in
the literature, the constraint $\Omega _{ANEC}<-\infty $ \ can
rule out well known macroscopic configurations (such as the one in
Eq.(21)) as being physically unrealistic.

\section*{Acknowledgments}

Our sincere thanks for administrative and technical support are
due to Guzel N. Kutdusova, Deputy Head of the Liason Office, BSPU,
Ufa, where part of the work was carried out. This work is
supported in part by the TWAS-UNESCO program of ICTP, Italy and
the Chinese Academy of Sciences, and in part by National Basic
Research Program of China under Grant No. 2003CB716300. NGM wishes
to acknowledge the financial support from Academy of Sciences of
the Republic of Bashkortostan Grant No. 3.2.1.5.

\end{document}